\title{Extreme mass ratio inspirals: perspectives for their detection}

\author{Stanislav Babak$^1$, Jonathan R. Gair$^2$, Robert H. Cole$^2$\footnote{Email: stba@aei.mpg.de, jrg23@cam.ac.uk} \\
        $^1$ Albert Einstein Institute,  Am Muehlenberg 1, D-14476 Golm, Germany\\
        $^2$ Institute of Astronomy, University of Cambridge, Cambridge, CB3 0HA, UK}

\date{\today}

\documentclass[11pt,english]{article}
\usepackage[]{graphicx}
\usepackage{amsmath, amssymb, amsfonts}
\usepackage{etoolbox}

\graphicspath{ {./Figs/} }

\newcommand{\rmd}{{\rm d}}

\begin{document}
\maketitle

\begin{abstract}
In this article we consider prospects for detecting extreme mass ratio inspirals (EMRIs)
using gravitational wave (GW) observations by a future space borne interferometric 
observatory eLISA. We start with a description of EMRI formation channels. Different 
formation scenarios lead to variations in the expected event rate and predict different 
distributions of the orbital parameters when the GW signal enters the eLISA sensitivity band. 
Then we will briefly overview the available theoretical models describing the GW signal from 
EMRIs and describe proposed methods for their detection.
\end{abstract}

\section{Introduction}
\label{intro}
Extreme mass ratio inspirals (EMRIs) arise following the capture of a small compact object (CO) --- a 
white dwarf, neutron star or stellar mass black hole --- by a massive 
black hole (MBH) in the centre of a galaxy. The astrophysical processes that lead to the formation of EMRIs are described in detail in section \ref{AstroEMRI}. 
The inspiralling CO loses energy and angular momentum through emission of gravitational radiation, and the initially wide and very eccentric orbit gradually shrinks and becomes more circular. EMRIs are among the most interesting gravitational wave (GW) sources that could be observed by the proposed eLISA detector. eLISA (evolving Laser Space Interferometer Antenna) is a space-based
gravitational wave detector which is scheduled for launch in 2034. It will be sensitive to GWs
in the frequency range $0.1 - 100$ milliHertz. Sources in this band include the mergers of massive black hole  binaries, which will be observable up to a redshift $z=20$, and numerous white dwarf binaries in the Milky Way, in addition to EMRIs. We will discuss the event rate and the expected  precision of parameter estimation for EMRI soures in section \ref{AstroEMRI}. 

During an EMRI, the CO typically spends $10^5 - 10^6$ orbital cycles in the eLISA band before plunging into the central MBH. 
We need to model the phase of GW signal from EMRIs with an accuracy of a fraction of a cycle in order to detect the signal and correctly extract the parameters of the binary system. This is a challenging problem, which has not yet been solved in full. Due to the extreme mass ratio, $m/M \sim 10^{-4} - 10^{-6}$, we can treat the 
problem perturbatively, considering the field of the CO and the emitted GWs as a small perturbation of the background spacetime of the central MBH. At the leading orders in mass ratio the internal structure of a CO is not important and so the CO is conventionally treated as a delta-function. As often happens in such an approximation, the self-field is divergent at the position of the CO, and requires proper treatment (regularization)~\cite{lrr-2011-7}. The resulting perturbation has the form of a tail expression, and depends on an integral over the entire past history of the CO's trajectory. In the limit that the mass ratio goes to zero, the motion is described by a geodesic. However, the mass of a CO is small but not zero and due to interaction of 
the self field of the particle with a background, the trajectory slowly deviates from a geodesic path \cite{GrallaWald2008}. This can be described effectively as the action of a  force (self-force) on the inspiraling object. In practice, the geodesic trajectory is used to compute the tail integral entering the self-force, and the resultant force is used to update the geodesic trajectory accordingly. In section~\ref{EMRIsWaveform}, we will summarize various ways to compute 
the GW signal from EMRIs and describe how the evolution of the orbital motion can be described using an osculating elements approach. The CO may also be spinning and this spin is coupled to the background curvature and alters the trajectory of the CO, forcing it to deviate from the corresponding geodesic of a non-spinning body. The trajectory of a spinning particle (in the limit of vanishing mass ratio) is described by the Mathisson-Papapetrou equation. Attaching a spin to a point particle is not uniquely defined, leaving a freedom to choose the dipole moment of a body (see \cite{MTW} for a description of spinning objects in the weak field approximation). This freedom manifests itself through the need to specify a spin supplementary condition (SSC) in order to obtain a unique solution to the equations of motion. In order to understand these complications, we consider in subsection~\ref{deSitter} the motion of a spinning particle in de Sitter space time. This space time possesses a 
non-trivial curvature but is still fully symmetric. For more details on the computation of the self-force and on the Matthisson-Papetrou equations we refer to other articles in this issue.

Last, but not least we want to consider the question of detectability of GW signals from EMRIs. 
The GWs generated by an EMRI system are characterized by 14 parameters: two masses $m, M$, the dimensionless spin of the MBH, $a$, and its orientation, $\theta_K, \phi_K$; six parameters describing the CO's position and velocity at some fiducial time or equivalently the instantaneous shape and phase of the orbit at that time (eccentricity $e$, inclination of the orbital plane to the spin of MBH $\iota$, semi-latus rectum $p$ and the initial phases $\phi_r, \phi_{\theta}, \varphi$ corresponding to the three coordinate degrees of freedom); the sky location of the source, $\theta, \phi$, and its luminosity distance, $D_L$. Many of these parameters are highly correlated. The GW signal comprises a superposition of orbital harmonics, with the number of harmonics and their relative strength strongly dependent on the eccentricity and binary orientation. The strength of the signal observed in the detector varies with time as eLISA moves around the sun (amplitude modulation) and the relative motion of the detector and the source induces a time-dependent Doppler modulation 
of the phase. The main challenge in detecting EMRIs is the multi-modality of the likelihood. The likelihood can be seen as a hyper-surface embedded in the 14-dimensional parameter space. It has multiple strong maxima and the main challenge is to find the highest (global) maximum. In section \ref{EMRIsDetection} we describe algorithms to do this which were successfully demonstrated on the Mock LISA data challenges \cite{Babak:2009cj}. 

Throughout this paper we use geometrical units $G=c=1$.

\section{Astrophysics of extreme mass ratio inspirals}
\label{AstroEMRI}

In this section we will consider possible channels leading to EMRI formation, the expected number of EMRI events that will be observed for eLISA and the likely accuracy with which eLISA will constrain their parameters. Then we will briefly summarize some of the potential impact of EMRI detections for astrophysics and fundamental physics.

\subsection{Formation of EMRIs}
The ``extreme mass ratio'' refers to the fact that the mass of the CO is of order of $1-10M_\odot$, while  the mass of the central (capturing) object is in the range $10^5 - 10^7M_\odot$. Current astrophysical observations indicate that massive compact objects of this kind are present in the nuclei of all sufficiently massive galaxies for which the central part can be resolved. The best example is the nucleus of the Milky Way, in which a few dozen bright O-B stars (so called S-stars) have been observed in Keplerian orbits around a central object with an estimated mass of $\sim 4 \times 10^6 M_{\odot}$. In addition, the compactness of this object suggests that it must be a massive black hole. 

These massive objects in the centres of galaxies are typically surrounded by clusters of stars. In the ``standard'' picture of EMRI formation, the stars are spherically distributed around the MBH (which should be approximately true for sufficiently large distances) and dense enough for efficient 2-body relaxation, i.e., mutual gravitational deflection and contact collisions. The timescale for this process, the relaxation time $t_{rlx}$, is defined as the time required to change the angular 
momentum of a star by an amount $J_c$, where $J_c$ is the angular momentum of a star on a circular orbit with the same semi-major axis. A smaller $t_{rlx}$ implies that stars can be more easily deflected on to very eccentric orbits with a small periapsis passage. If a CO object on an initially wide orbit is perturbed onto such a trajectory, it will lose energy to GW bursts emitted near periapsis ($r_p$) and its orbit will gradually shrink. While the semi-major axis is very large, the CO can still efficiently interact with other stars at the apoapsis and could be either deflected onto a plunging orbit with $r_p < 8M$ or onto a wide orbit which does not emit appreciable GW radiation. To become an observable EMRI, the CO must remain on the highly eccentric orbit until its period becomes smaller than $\sim 10^3-10^4$ s, at which point it is continually radiating GWs in the eLISA sensitivity band. While we will be primarily interested in such EMRIs here, the bursts of GWs produced during periapsis passages in the early stages of the process could also be potentially detected by eLISA if the event is in the nucleus of nearby galaxies~\cite{BerryGair2010}. 
 
When the stars interact gravitationally, they tend to divide the kinetic energy equally and, while equipartition is not reached in practice, this process causes more massive objects to sink deeper in the potential well of the MBH. This process is called mass segregation. As a result we expect stellar mass black holes to  form a steep power-law density cusp around the MBH $n(r) \sim r^{-\alpha}$ with $\alpha \simeq 1.7-2.0$, which dominates for $r<0.1$pc. The lighter stellar species form shallower density profiles with  $\alpha \simeq 1.3-1.5$~\cite{AlexanderHopman2009ApJ}. The relaxation time is inverse proportional to the density of the CO and it should therefore be smaller for the stellar mass black holes. 
  
In order for an object to become an EMRI, it should efficiently dissipate energy through GW emission, and have a sufficiently low probability to be deflected onto a different orbit. This condition implies that the time scale for orbital decay by GW emission, $t_{GW}$, should be smaller than $(1-e)t_{rlx}$, where $e$ is orbital eccentricity. Once the orbital period reaches $P < 10^4$s, the CO completely decouples from the cusp, which happens for orbits with semi-major axis $a_{EMRI} \sim 0.05$pc.

For typical orbits around an MBH, the number of stars enclosed by the orbit is rather small and so
the gravitational potential created by the ``field" stars is not a smooth symmetric function.
This gives rise to a torque acting on a CO on an orbit with semi-major axis $a_{CO}$ of $\tau \sim \sqrt{N} m_{*}/a_{CO}$, where $N$ is a number of field stars with mass $m_*$ inside the CO orbit. If the 
precession of the CO orbit is slow compared to the timescale over which the distribution of field stars changes significantly, the CO experiences a nearly constant torque over some time. This mechanism, known as resonant relaxation, changes the angular momentum of the CO, but not its energy. The characteristic time scale associated with resonant relaxation, $t_{RR}$, is significantly smaller than $t_{rlx}$ and so this process can significantly boost the EMRI event rate. Resonant relaxation plays an important role for orbits with $a_{CO} \le a_{EMRI}$ \cite{HopmanAlexander2006}, \cite{HopmanAlexander2006ApJ}, \cite{Eilon:2008ht}, \cite{AlexanderBook}. However, for COs on eccentric orbits with small perhaps radii, the relativistic (GR) precession can be very high, which effectively destroys the resonant relaxation effect. The point at which this occurs is known as the ``Schwarzschild barrier'' \cite{Merritt:2011ve}. The existence of this barrier means that resonant relaxation is not as effective at boosting EMRI rates as one might first think, although if the MBH has significant spin then the impact of the ``Schwarzschild barrier'' is somewhat diminished due to the lower value of the plunge periapsis for prograde orbits~\cite{Brem:2012et}. In this case, COs that would normally be considered as plunging and hence undetectable around a Schwarzschild MBH actually perform many cycles in the eLISA band and may contribute significantly to the event rate~\cite{Amaro-Seoane11032013}.

In this picture, the critical thing for having a high EMRI rate is to have compact objects in the ``loss-cone'' (orbits with impact parameter sufficiently small that they can be captured or tidally disrupted by the MBH). Several channels have been suggested that can replenish the loss-cone and thereby significantly boost the EMRI rate, including triaxiality of the potential (non-spherical galactic nuclei) \cite{Merritt:2010qv} or the presence of massive perturbers (such as intermediate mass BHs, and/or molecular clouds) in the vicinity of the orbits \cite{Perets:2006bz}.

The complex dynamics of this standard capture scenario for EMRI formation means that the astrophysical event rates are very uncertain. To estimate event rates we will use a current best guess of $400$Gyr$^{-1}$ for Milky Way-like black holes, dominated by EMRIs in which the CO is a black hole. This rate is taken from~\cite{Hopman:2009gd}.

As well as this standard mechanism for EMRI formation, there are two other plausible channels. 

\emph{Tidal binary disruption.} It is possible that within the radius of influence of a MBH there is a binary 
fraction of at least a few percent \cite{ColemanMiller:2005rm}. If a binary approaches the MBH it can be tidally disrupted and, if this happens, one star is ejected at very high velocity while the other star becomes tightly bound to the MBH. The captured CO is expected to end up on an orbit with a semi-major axis of a few hundred AU and a pericentre distance of a few to tens of AU, implying that it will circularise by the time it enters the 
eLISA frequency band~\cite{ColemanMiller:2005rm, Hopman:2009gz}. This is a distinct feature of this formation channel, since in the standard scenario we expect the EMRIs to have a significant residual eccentricity even at plunge \cite{AmaroSeoane:2007aw}, $e_{pl}\sim 0.1-0.3$. We observe in the Milky Way so called ``hyper-velocity'' stars  \cite{Brown:2008ck}. which are moving away from the galactic centre with large velocities. The best current explanation for the presence of short-lived S-stars in the vicinity of the Milky Way MBH is that they came there following the tidal disruption of binaries, while the observed hyper-velocity stars are the thrown away companions \cite{Svensson:2007yx}.

\emph{Formation of stellar remnants in a disk.} 
Observation of active galactic nuclei suggests the presence of a circum-nuclear gaseous disk accreting onto the MBH. If the disk is thick and sufficiently massive, the outer part could fragment and form stars. If migration through the disk is sufficiently slow, stars formed in this way could evolve to form compact object remnants (neutron star or black hole) which subsequently spiral into the MBH as an EMRI in the equatorial plane (the accretion disk, at least its inner parts, is expected to be aligned with the MBH's equatorial plane \cite{Levin:2006uc}). The interaction with the gas is also likely to keep the orbit of a CO close to circular, so the distinct feature of this channel of EMRI formation is a circular orbit in the black hole equatorial plane.

By measuring the orbital parameters we will be able to say which of these three channels provides the most likely explanation for how the EMRI was formed. For more details on the dynamics of galactic nuclei we refer the reader to the comprehensive review~\cite{AmaroLR}.

\subsection{Expected event rate estimation}
In this section we follow \cite{Gair:2008bx} and briefly outline how the expected event rate of EMRIs observed by eLISA can be estimated. To make this estimation we require an intrinsic event rate $\mathcal{R}(M, a, \mu$), where $M$ is the mass of the MBH, $a$ its spin in units of $M$ and $\mu = m/M$ is the mass ratio. The intrinsic event rate tells us how often EMRIs are formed (i.e., how often they enter the eLISA sensitivity band) per galaxy hosting a MBH with parameters $M, a$. The mass ratio parameter tells us the nature of a CO, i.e., whether it is a stellar mass BH, neutron star or white dwarf. As discussed in the previous subsection, due to mass segregation we expect stellar mass BHs to be the most likely candidate for EMRIs, so we choose a canonical value for the CO mass of $m=10 M_{\odot}$. We will normalize the mass of the MBH by the mass expected for a Milky Way type galaxy $M_{MW} ~ 3\times 10^6 M_{\odot}$. So far we do not have information about the distribution of the spin of MBHs of this mass. X-ray observations of some active galactic nuclei  provide information about the spin of accreting MBHs in the centre, but those black holes are of higher mass  $>10^7 M_{\odot}$ and embedded in the gaseous circum-nuclear disk. In addition all present estimations of the spin are heavily model dependent and could vary significantly depending on the underlying assumptions~\cite{Brenneman:2013oba}. Therefore, here we assume a uniform distribution of the spin within its physical range $a \in (-1, 1)$. The estimation of the intrinsic event rate is a very challenging task, as described above and in more detail in~\cite{AmaroLR}, which depends quite heavily  on the  underlying assumptions about the efficiency of mass segregation, the relative importance of different EMRI formation channels and the interplay between resonant relaxation and the ``Schwarzschild barrier''.  Here we adopt the estimate derived in \cite{Hopman:2009gd} which for stellar mass BHs is 
  \begin{equation}
  \mathcal{R}  = 400 Gyr^{-1}\left(\frac{M}{3\times 10^6M_{\odot}}\right)^{\beta}
  \label{Eq:beta}
  \end{equation}
  where $\beta \approx 0.19$.
  
If the duration of EMRI signals was significantly shorter than the observation time, then the observed event rate would be determined by computing the distance at which the signal-to-noise ratio (SNR) equals some detection threshold $\rho_{thr}$ and then multiplying the rate per unit volume by the volume contained by that distance, assuming a uniform distribution of EMRIs in the local Universe. However, EMRIs are long-lived, and the SNR can be accumulated for as much of the inspiral as coincides with the eLISA observation. Fixing all the parameters of the EMRI, we can compute the SNR as a function of the time left to plunge, $t_{pl}$. As we increase $t_{pl}$ from zero, the SNR 
first increases, then reaches a maximum before starting to decrease. There is a decrease of SNR for large $t_{pl}$ because the finite observation time means that we are ultimately only observing systems that are rather wide, with not very efficient GW emission, and with emission primarily at low frequencies where acceleration noise rises rather steeply. This means that if an EMRI is at all detectable, the SNR as a function of $t_{pl}$ intersects the line SNR$=\rho_{thr}$ at two times, $t_{early},\; t_{late}$, and we can define the EMRI observable lifetime as $\tau(\lambda_i) = t_{late}(\lambda_i) - t_{early}(\lambda_i)$, where $\lambda_i$ corresponds to all the other parameters of the EMRI (besides $t_{pl}$) which we have fixed. If EMRIs plunge at a rate $\mathcal{R}$ per year in a particular galaxy, then $\tau \mathcal{R}$ gives the expected number of events from that galaxy (after appropriate averaging over parameters $\lambda_i$). Among all parameters describing the 
EMRI system the most important are $M, a$, and we denote the remaining parameters as $\hat{\lambda}_i$. 
We define $\mathcal{N}(M,a,z)dMda$ as the number of MBHs per comoving volume with mass
$M\in [M, M+dM]$, spin $a\in [a, a+da]$, and at redshift $z$. We make 
two further assumptions (i) that the mass and spin distributions are independent; and (ii) that the distribution of MBH mass and spin are independent of redshift. The first assumption reflects our level of ignorance, and the second assumption is reasonable given how far we can observe EMRIs (with eLISA we will able to see EMRIs up to $z_{\rm max}\approx 0.7$). In this range we can ignore the evolution of masses and spins with $z$. Under these assumptions $\mathcal{N}(M,a,z)dMda = (dn/d\ln{M})(M) d\ln{M} p(a) da$, where $p(a)$ is the probability distribution function for the spin $\int p(a) da = 1$. As described above, we assume this is uniform in our calculations, but we keep it here in the equation for completeness. 

The expected event rate is then
\begin{eqnarray}
  N_{eLISA} &=& \left<
  \int_{z=0}^{\infty}dz\int_{M_{low}}^{M_{high}} d\ln{M}\int_{a=-1}^{1} da 
  \mathcal{R}(M,a)\tau(M,a,z, \hat{\lambda}_i) \right. \nonumber \\
  & & \left.
   \frac{dn}{d\ln{M}}(M) p(a)
  \frac{dV_c}{dz} 
  \right>_{\hat{\lambda}_i}
\end{eqnarray}
Here $(dV_c/dz)dz$ is the comoving volume in the redshift range $[z, z+dz]$. The triangular brackets 
denote the averaging over other EMRI parameters $\hat{\lambda}_i$. We note that in practice the intrinsic event rate could also depend on some parameters from the set $\hat{\lambda}_i$ (depending on the channel of EMRI formation). The mass function ($dn/d\ln{M}$) can be deduced from measured galaxy luminosity functions using the observed $L-\sigma, M-\sigma$ correlations. In the range of interest to eLISA, this functions approximately flat~\cite{Shankar:2007zg}, so we adopt 
\begin{equation}
 \frac{dn}{d\ln{M}} = n_0 \left(\frac{M}{3\times 10^6M_{\odot}}\right)^{\alpha} 
 \label{Eq:mf}
\end{equation} 
with canonical values $n_0 = 0.002\; \rm{Mpc}^{-3}, \;\; \alpha =0$. 
If we assume these canonical values, with $\beta =0.19$, a mission duration of 2 years and a detection threshold of $\rho_{thr}=20$, we estimate that eLISA would observe 25 to 50 events in two years \cite{AmaroSeoane:2012je, AmaroSeoane:2012km}. This spread in the predicted number of events comes from uncertainties in the waveform model and system parameters, but a much larger uncertainty, which is not taken into account here, arises from the uncertainty in the true value of $\mathcal{R}$.  


\subsection{Science return from observing GW signals from EMRIs}

\begin{figure}[ht]
\center{
\includegraphics[height=0.5\textheight,keepaspectratio=true]{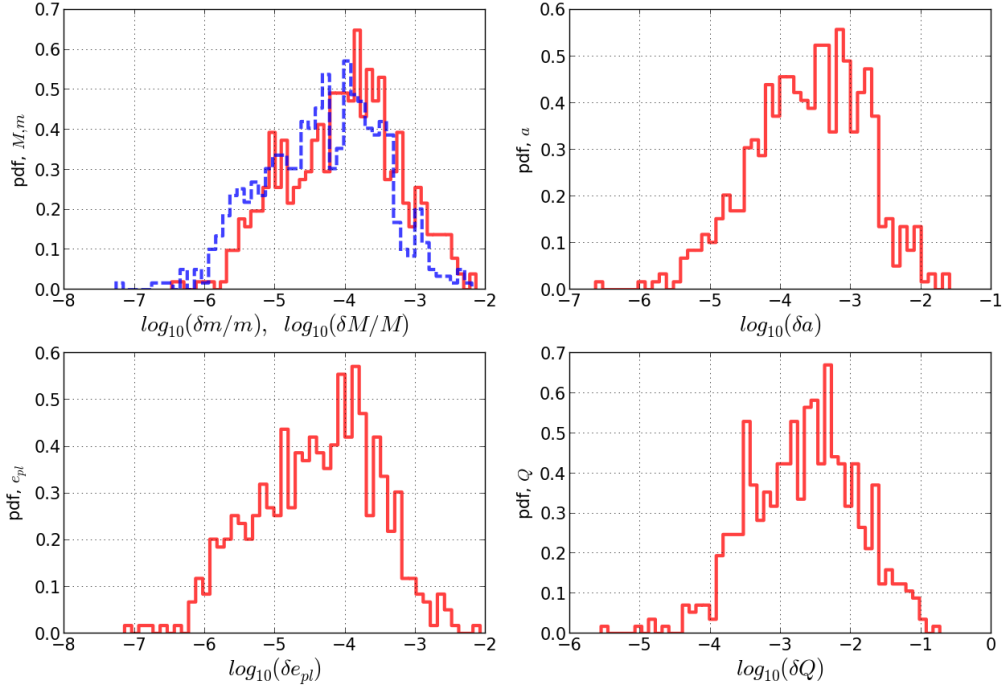}
}
\caption{Expected precision of parameter estimation from observed EMRI events, computed using the Fisher information matrix: MBH mass ($M$, dashed line),
CO mass ($m$, solid line), MBH spin ($a$), orbital eccentricity before plunge ($e_{pl}$) and deviation of the MBH quadrupole moment from the Kerr value ($Q$).}
\label{Fig:EMRIparams}
\end{figure}

Detection of EMRIs and measurement of their parameters provides unique astrophysical data which cannot be obtained by any other means. We expect to be able to learn information about stellar populations in the centre of the Milky Way in the future by observing pulsars in the nuclear stellar cluster region using the SKA \cite{Liu:2011ae}. Inferring similar properties of other galaxies through observations of EMRIs will also us to compare the nucleus of the Milky Way with nuclei of other galaxies. The number of observed EMRI events and the mass distribution of the COs will tell us 
about the physics of mass segregation, the masses and spins of stellar mass compact objects and about the steepness of the stellar cusps in the centres of galaxies. In addition, EMRI observations will provide precise measurements of MBH masses and spins in a new mass range. EMRIs will probe galaxies containing black holes with masses $10^5 - 10^7 M_{\odot}$, and such galaxies tend to be of lower mass and not particularly luminous in the electromagnetic spectrum. Extracting information about the nuclei of those galaxies is therefore very challenging, if not impossible, using electromagnetic observations and eLISA therefore has tremendous potential to inform us about these systems. Observations show that the masses of black holes in galactic nuclei correlate with the mass, luminosity and the stellar velocity dispersion of their host galaxy~\cite{Gultekin:2009qn}. These correlations imply that black holes evolve along with their hosts throughout cosmic time, but it is not yet known if this coevolution extends down to the lowest galaxy and black hole masses, since those systems may have differences in the accretion properties \cite{Mathur:2005xd}, dynamical effects \cite{Volonteri:2007tu}, or cosmic bias
\cite{Volonteri:2009vh}. eLISA observations of EMRIs will significantly improve our knowledge of the 
MBH mass function (e.g., inferring the parameter $\alpha$ in eqn.~[\ref{Eq:mf}]), as well as allowing us to measure the 
intrinsic event rate (for example constraining the parameter $\beta$ in eqn.~[\ref{Eq:beta}]), determine the relative importance of different channels of EMRI formation and measure the spatial distribution (relative to the MBH) of different types of CO. This is made possible by the 
ultra-precise determination of EMRI parameters with GW observations. In Figure~\ref{Fig:EMRIparams}
we show how accurately we expect to measure the most important parameters: MBH mass ($M$) and spin ($a$), 
CO mass ($m$), orbital eccentricity just before the plunge (end of inspiral) ($e_{pl}$). The last parameter, $\Delta Q$,
is a possible deviation in the MBH quadrupole moment away from the Kerr value, which will be discussed at the end of this subsection.

In addition an observation of an EMRI will allow us to determine the luminosity distance to the source ($D_L$) with an accuracy of 
$\le 1$\% and to localize the source on the sky to about $0.2$ square degrees. Such a fantastic accuracy is 
achieved because the source is long lived --- the CO spends $10^4-10^6$ cycles in the close vicinity of a MBH. 
Using matched filtering we will be able to determine the phase of an EMRI to an accuracy of half 
a cycle, a fractional phase accuracy of $10^{-6}$--$10^{-4}$. All information about the binary system is encoded in the GW phase and so we can expect to make measurements of the intrinsic parameters to this same fractional accuracy. Measurements of the extrinsic parameters, such as sky localisation, are not as precise since these measurements come not from the phase but from the modulation of the GW signal (in amplitude and in phase) caused by eLISA's orbital motion. 

These precise phase measurements mean we can also use EMRIs to test the "no-hair" theorem: if the central massive
compact object is indeed described by the Kerr metric, as general relativity predicts. The spacetime outside a stationary, axisymmetric object is fully determined by its mass, $M_l$, and current, $S_l$, multipole moments. Since these
moments fully characterise the spacetime, the orbits of the smaller object and the gravitational waves it emits are
determined by these multipole moments. The emitted GWs therefore encode a map of the spacetime structure and by observing these gravitational waves with eLISA we
can precisely characterise the multipole structure of the central object. Extracting the moments from the EMRI
waves is analogous to geodesy. If the central object is a Kerr black hole, then all multipole moments are determined 
by its mass and spin (``no-hair'' theorem):
$$
M_l + iS_l = (ia)^l M^{l+1}
$$
If we can measure the first three moments we can therefore check whether the central object is consistent with being a Kerr black hole. Figure~\ref{Fig:EMRIparams} shows that we should be able to measure a deviation in the mass quadrupole moment from the Kerr value, $Q = |M_2 - M_2^{Kerr}|$, to a precision of $\delta Q \approx (10^{-2} - 10^{-3})M^3$.
EMRIs could therefore also serve as laboratories for testing fundamental physics. For more discussion on this topic we refer to \cite{AmaroSeoane:2012km,  Babak:2010ej}.

\section{Modelling the GW signal from EMRIs}
\label{EMRIsWaveform}

For detection of EMRIs we will utilize matched filtering, this technique assumes that we can model the 
GW signal and then cross-correlate it with the data. The EMRI signal depends on 14 parameters 
(actually on 17 if we take into account the spin of CO), which we do not know a priori and need to infer from the measured data. To do this, we must generate many signals from a given  model (templates) across the full, 14-D, parameter space to find the parameters that best fit the data
(this set of parameters, which maximize the likelihood, are called ``maximum likelihood estimators" of those 
parameters). We will describe the search procedure in detail in the next section. 

The presence of noise in the data stream causes the best-fit parameters to differ from the true parameters of the GW signal. The size of this difference can be estimated using the Fisher information matrix, as shown in Fig.~\ref{Fig:EMRIparams}. If the data is analysed using an inaccurate model there will also by systematic errors in the parameter estimates, which could be larger than the statistical errors from detector noise. It is therefore important to accurately model the GW signal coming from EMRIs, to ensure reliable estimation of parameters and improve the detectability, since a mismatch between the signal and template will cause a drop in the SNR and decrease in the observed volume by $(SNR/SNR_{optimal})^3$.

In this section we will describe currently available models for EMRI signal and discuss their 
effectualness (if they are able to recover the optimal SNR) and faithfulness (if the systematic errors 
in parameter estimation is below the statistical errors due to presence of the noise).
Note that effectualness does not imply faithfulness: a model could recover a significant 
fraction of the SNR with a large systematic bias in the parameters. In other words, 
the shift in the parameters from the true values could (partially) compensate for inaccuracies in the model.

\subsection{Waveform inventory}
\label{Waveforms}

Unlike the inspiral of a comparable mass binary, the merger (here we call it plunge) of a CO with 
MBH and subsequent ring-down are suppressed by a factor of the mass-ratio and are therefore not observable by eLISA.
We therefore only need to model the inspiral part of the signal up to a plunge. However, for the whole of the inspiral observable by eLISA, the CO is orbiting in the strong-field region close to the MBH and moving at ultra-relativistic speeds. This makes modelling an EMRI signal somewhat different from modelling GWs from a binary 
of two nearly-equal mass MBHs. Here we briefly outline some of the currently available models for GW signals from EMRIs. 
More detailed description can be found in other papers in this volume.

\emph{Post-newtonian expansion}. The post-newtonian approach describes the GW signal as an expansion in
velocity $v$. As mentioned above the CO in EMRI systems are fast moving and 
spend $10^4-10^6$ cycles in a regime where $v$ is large. The EOB approach \cite{Buonanno:1998gg, Damour:2009ic} is the most suitable for modelling EMRIs by construction (the conservative dynamics reduces to the test-mass in the limit $m/M \to 0$), however
the dissipative part (fluxes) are needed to a very high post-newtonian order, which is not currently known. In addition, the analytic expressions for the fluxes are known only for nearly circular and nearly equatorial orbits, while we expect EMRI orbits to be both eccentric and inclined~\cite{Mino:1997bx}.

\emph{Analytic "kludge" waveforms}.  This model was introduced primarily to study  
detection rates and parameter estimation for EMRIs~\cite{Barack:2003fp}. The main advantage of these waveforms 
is that they are fast to generate, so they are suitable for large Monte-Carlo simulations, and they were extensively 
used to develop detection algorithms (see section \ref{EMRIsDetection}). 
This model is an extension of the work by Peters and Mathew \cite{Peters:1963ux}, it represents emission from a
CO in Keplerian orbit augmented by imposing (post-newtonian) relativistic precession of the orbital plane 
and the direction to perihelion. The dissipative evolution is taken from post-newtonian calculations. This model is not 
particularly accurate but it captures the main physical processes occurring in EMRIs.

\emph{Numerical "kludge" waveform, or semi-relativistic model.} 
The idea of the numerical kludge waveforms is to combine an exact particle trajectory (up to inaccuracies in
the phase space trajectory and conservative radiation reaction terms) with an approximate expression for the GW
emission. By including the particle dynamics accurately, we hope to capture the main features of the waveform, even if we are using an approximation for the waveform construction.
The idea was introduced in  \cite{Ruffini:1981af, Tanaka:1993pu} and was further evolved with some modifications  in \cite{Babak:2006uv, Sopuerta:2011te}. 

The procedure to compute a numerical kludge waveform has two stages. Firstly, a phase-space inspiral trajectory is constructed, i.e., the sequence of geodesics that an inspiral passes through, by integrating prescriptions for the evolution of the six constants of the motion (energy, angular momentum, Carter constant and three initial phases). Initial work has used post-newtonian expressions (augmented by some consistency corrections and by fitting to solutions of the Teukolsky equations) to evolve these constants. This inspiralling trajectory is computed numerically thus the name ``numerical kludge''. Once the trajectory has been constructed a waveform is generated by identifying the Boyer Lindquist coordinates along the trajectory with spherical-polar coordinates in a flat space time and applying weak-field GW emission formulae, in particular the quadrupole-octupole approximation:
\begin{equation}
    \bar{h}^{jk} = \frac{2}{D_L}\left( \ddot{I}^{jk} - 2n_i \ddot{S}^{ijk} +
    n_i \dddot{M}^{ijk} \right)|_{t' = t-D_L},
\end{equation}	
where $I^{ik}, M^{ijk}$ are the mass quadrupole and octupole moments and $S^{ijk}$ is the current quadrupole moment of the binary system, 
$n^i$ is a unit vector pointing from MBH to the position of a CO, and overdots denote time derivatives. These waveforms are somewhat slower to generate as compared with the analytic kludge due to the numerical integration of the orbital trajectory, but it is far more faithful up to the last month or less (semilatus rectum $p\approx 6M$) before the plunge.
 
 \emph{Adiabatic inspirals based on Teukolsky formalism}.  The very first framework for black hole perturbation theory in a Kerr background was the Teukolsky formalism \cite{Teukolsky:1973ha}, which encapsulates all gravitational radiative degrees of freedom in a single "master" wave equation  
(the ``Teukolsky equation'') for the Weyl scalars, $\Psi_0$ and  $\Psi_4$. A key feature of this equation is that it admits separation of variables
in the frequency domain, which effectively reduces it to a pair of ordinary differential equations. 
The Teukolsky equation has been solved in the frequency domain \cite{Drasco:2005kz} and in the time domain \cite{Martel:2003jj},
but both approaches  assume the orbit that acts as the source of the perturbation is a geodesic. The rate of change in energy, angular momentum, Carter constant (averaged over several orbits) are evaluated from the gravitational wave field and then used to update the parameters of the geodesic in an adiabatic manner. 
This procedure misses the evolution of the other constants of motion (initial positions) as well as making the
adiabatic assumption. As a result, the waveforms are not accurate on a very long time scale, but 
they are the most faithful model on time scales $\sim M^2/m$. 

\emph{Self-force waveforms}. An accurate description of the self-force and its derivation is given 
in other papers in this volume, so we only briefly mention it here. 
As mentioned above, the extreme mass ratio in an EMRI system allows the waveform to be determined using
perturbation theory. The inspiralling object can be regarded as a small perturbation on the background spacetime of
the central black hole, except very close to the small object. In the vicinity of the small object, the 
spacetime can be
regarded as a Schwarzschild BH moving under the influence of an external tidal field due to the MBH. 
Matching these two regimes allows one to obtain an expression for the self-force acting on the CO. 
The self force can be seen to arise as a result of the interaction of the self field of the CO with the 
non-flat background geometry, which causes the lines of force to be bent and act back on the CO. 
The self-force can be conventionally split into two parts: non-time symmetric (dissipative) and 
time-symmetric (conservative). The former part causes the inspiral and dominates
while the latter part can be eliminated by a redefinition of the orbital frequencies at each instance, which means it is effectively second-order in mass ratio. 
The adiabatic Teukolsky based waveforms take into account only the dissipative part of the self force,
neglecting the conservative part, which defines the domain of its validity. 
The self force is computed assuming the CO is moving on a geodesic, then it is used to adjust the geodesic (inspiral) 
before the self-force is recomputed again. The computation of the self force is somewhat complicated 
as it treats the CO as a delta function in the background spacetime, which requires mathematical 
apparatus for regularization of some divergent integrals. It is possible to subtract the singular part 
from the field equations (by finding a singular solution valid in the vicinity of the CO) and the resulting 
equations are manifestly regular and contain on the right hand side a smooth effective source
\cite{Vega:2011wf}, which allows the field equations to be coupled to the equations of motion 
and integrated. This procedure can be written for a scalar field (representing a CO carrying a scalar charge and ignoring the gravitational part of the self-force) as 
\cite{Diener:2011cc}
\begin{eqnarray}
   {(\Phi^r)_{;\alpha}}^{;\alpha} = S(x; z(\tau), u(\tau))\\
	\frac{Du^{\alpha}}{d\tau} = \frac{q}{m(\tau)} (g^{\alpha\beta} + u^{\alpha}u^{\beta})
	\bigtriangledown_{\beta}\Phi^r \\
	\frac{dm}{d\tau} = - q u^{\beta}\bigtriangledown_{\beta}\Phi^r, 
\end{eqnarray} 
where $\Phi$ is the scalar field, $q$ is the scalar charge, $m$ the mass of the CO, $S$ is an effective source term
and $u^{\alpha}$ is the CO four-velocity. Greek indices are being used to indicate space-time components, and a semicolon 
denotes a covariant derivative with respect to the background spacetime, $g_{\alpha\beta}$.
A similar procedure can be applied to the gravitational field. So far only the self-force waveform 
 for a Schwarzschild background has been computed, but recent progress has been rapid and so we expect 
the extension to Kerr to be completed within a few years.

\emph{Numerical relativity waveforms}. The ultimate goal would be to compute EMRI waveforms using numerical integration of the 
full GR field equations. State-of-the-art techniques have enabled the computation of waveforms for the last 
20-50 cycles of the inspiral, merger and ring-down of comparable mass ratio binaries. The simulation of an
EMRI requires the computation of a few orders of magnitude more cycles, plus the resolution of two very different spatial scales. This is far beyond the capability of current 
computational resources and techniques. In addition, the time step for explicit numerical integration 
is set by the smallest characteristic scale in the problem, which is the mass of the CO in this case. 
Numerical waveforms will be very useful for the calibration of current calculations based on perturbation techniques, 
but new numerical methods will have to be developed to handle EMRIs.

\subsection{Evolving perturbed geodesic motion}

In this subsection we will focus on how we can compute the evolution of the orbit. The orbital evolution is the key ingredient
for creating numerical kludge waveforms and waveforms based on the self force. In fact this is the same problem, 
the main difference is in how the waveform is computed from the orbital trajectory. 
To compute the orbital evolution we must solve the forced geodesic equation:
\begin{equation}
  u^{\beta}{u_{\beta}}^{;\alpha} =  a^{\alpha},
  \label{Eq:forcedmot}
\end{equation}
where $a^{\alpha}$ is the 4-acceleration. The acceleration is essentially the self-force, but the method we will 
describe here for solving this equation is also applicable to the case  where $a^{\alpha}$ represents some other kind of external perturbation. 
This perturbation could be caused by a second (intermediate) MBH (if the CO in the EMRI is inspiralling into an MBH that is in a wide MBH binary), a molecular cloud or disc, another star or compact objector basically anything that can cause a slow modification of the geodesic orbit. Here we assume that 
the acceleration has been derived in some other way and are only interested in the effect it has on the inspiral trajectory. 

The rest of this subsection summarizes results described in more detail in~\cite{Gair:2010iv}. We use an osculating elements approach 
to evolve eqn.~(\ref{Eq:forcedmot}). If the perturbing force is small~\footnote{In fact this formalism does not assume the force is small --- there is a unique geodesic passing through any given point with a particular velocity and so any trajectory can be described as an osculating geodesic. However, the approach is most useful when the force is small since then the trajectory remains almost geodesic and parameterising it in terms of instantaneous geodesic motion is useful.}, we can represent the perturbed
trajectory at each instant by the unique geodesic passing through the same position with the same velocity and see the orbital
evolution as a slow variation of the constants of these instantaneously-tangent geodesics. A general geodesic in Kerr spacetime 
is described by eight constants of motion:  $J = \{m,E, L_z, Q, \psi_0, \chi_0, \phi_0, t_0 \}$, however two of them 
(CO mass and initial time $m, t_0$) are not truly dynamical, so we will work with the remaining 6:
orbital energy ($E$), orbital angular momentum projected onto the spin of the MBH ($J_z$), Carter constant ($Q$) and
three initial phases ($\psi_0, \chi_0, \phi_0$) describing the initial position of the CO on the orbit 
in $r, \theta$ and $\phi$ respectively. The osculating element for of the equation of motion $\ddot{\bf r} = {\bf f}_{geo} + \delta{\bf f}$, is  
	\begin{eqnarray}
			z^{\alpha}(\tau) = z^{\alpha}_g(J^A(\tau), \tau), \;\;\; \to
			\frac{\partial z^{\alpha}_g}{\partial J^A} \frac{\partial J^A}{\partial \tau} = 0\\
			\frac{\partial z^{\alpha}}{\partial \tau} = \frac{\partial z_g^{\alpha}}{\partial \tau}
			(J^A(\tau), \tau), \;\;\; \to  \frac{\partial \dot{z}^{\alpha}_g}{\partial J^{A}}
			\frac{\partial J^A}{\partial\tau} = \delta f^{\alpha}.
			\label{Eq:OE}
		\end{eqnarray}	
 The first set of equations describes a ``geodesic" motion with slowly changing orbital ``constants'', and the 
 second set gives us the evolution of the orbital ``constants'' as a function of the perturbing force.
 
 The  advantage of using the osculating elements approach is that we can use an adiabatic approximation
 (or, more  generally, a two-time-scale expansion \cite{Hinderer:2008dm})
 to evolve EMRIs, for which the radiation reaction time scale is much longer than the orbital time scale,  allowing us to more easily study secular effects.

The osculating elements approach was first used in \cite{Pound:2007th} to study Eq.~\ref{Eq:forcedmot}
in  Schwarzschild background, and was extended to Kerr in \cite{Gair:2010iv}. The authors 
in \cite{Gair:2010iv} wrote the osculating element equations on two different forms, using the Kinnersley tetrad or ``Hughes'' variables (i.e., in
terms of the orbital constants and the total phase variables \cite{Drasco:2005kz}). In both cases, the
appearance of an apparent divergence in the osculating equations of motion at turning points is avoided.
The  techniques were applied to a toy problem in which an EMRI was evolving under the influence of a perturbing force due to drag from surrounding material. This ``gas-drag'' force
was taken to be proportional to the velocity of the inspiralling compact object. The two different approaches were shown to give identical results, and the comparison of  the exact and adiabatic solutions to the problem identified the domain of validity of the adiabatic approach. Although the gas-drag problem was considered only to illustrate the methods, it yielded interesting results. In particular, it was found that the influence of the drag force was to drive the inspiral of the object, but also to increase the eccentricity of the orbit and decrease the orbital inclination. A gravitational wave driven inspiral would tend to show a decreasing eccentricity and so these two types of perturbing force would be distinguishable in an EMRI observation.

Osculating elements were also used to generate inspirals in a Schwarzschild background under the influence of the gravitational self-force in~\cite{2012PhRvD..85f1501W}. The formalism developed for Kerr inspirals in~\cite{Gair:2010iv} has not been used for any other studies so far, but this will be done once suitable models for perturbing forces are available. Another type of orbital perturbation, which can also be interpreted in terms of a perturbing force acting on a geodesic, is the influence of the spin of the CO on the trajectory. This will be discussed in detail in the next subsection.

\subsection{Spinning particle in de Sitter space-time}
\label{deSitter}

In this subsection we will consider a spinning CO. There are several contributions to this 
proceedings which describe the motion of a spinning body in a given background in 
great detail. Here we will give only a brief summary, then show how we can 
formulate the motion in terms of the osculating elements approach described in the last subsection. To understand the 
motion of a spinning CO in the MBH spacetime, we will first consider a simpler problem. We will describe analytically the motion of a spinning test body in de Sitter spacetime. 

The motion of a test mass in an arbitrary spacetime is governed by the Mathisson-Papapetrou 
equations
\begin{eqnarray}
	D_{\tau} p^{\alpha} = -\frac1{2} {R_{\mu\nu\beta}}^{\alpha} u^{\beta}S^{\mu\nu}\label{eq:MP1}\\
	D_{\tau} S^{\alpha\beta} = 2 p^{[\alpha}u^{\beta]}.
	\label{Eq:MP}
\end{eqnarray}	
The first complication is that the 4-momentum $p^{\mu}$ and 4-velocity $u^{\mu}$ are not parallel
\begin{equation}  
p^{\alpha} = m u^{\alpha} + u_{\beta} D_{\tau}S^{\alpha\beta}.
\label{Eq:mom_vel}
\end{equation}
Here $D_{\tau}$ denotes a covariant derivative with respect to the proper time,
square brackets denote the anti-symmetric part, 
${R_{\mu\nu\beta}}^{\alpha}$ is the Riemann tensor of the background space time and 
 $S^{\mu\nu} = - S^{\nu\mu}$ is the spin tensor. The difference between $p^\alpha$ and $u^\alpha$ means that there is an ambiguity in what we call the
 mass --- we can define this as $m=p^{\alpha}u_\alpha$ or $M^2 = p^{\alpha}p_{\alpha}$. The second complication is that 
 there is not a sufficient number of equations to determine all of the unknowns. In order to 
 close the system we need to introduce an additional ``spin supplementary condition'' (SSC). 
 There is an arbitrariness in choosing the SSC, which is usually attributed to how we 
 choose the representative word line of a test mass (this is equivalent 
 to choosing a dipole moment of a spinning CO). The main reason that the SSC is needed is that there is an ambiguity in 
 the definition of the spin tensor for a point mass. The point mass is 
 an approximation of an extended body (for which the spin tensor is well defined) 
 when the size is much less than the radius of curvature of the background spacetime. 
 The most common SSCs are 
\begin{equation}
(i)\; \; p_{\alpha}S^{\alpha\beta} = 0, \; (ii)\;\; u_{\alpha} S^{\alpha\beta} = 0,\;
(iii)\;\; w_{\alpha}S^{\alpha\beta} = 0
\end{equation}
SSC (i) is usually referred to as the Tulczyjew condition \cite{Tulczyjew}, 
(ii) is the Frenkel-Pirani condition \cite{Frenkel:1926zz, Pirani:1956tn} and (iii) was first introduced in 
\cite{Kyrian:2007zz} and is referred to as the $w$-condition. 

As mentioned above we want to write the Mathisson-Pappetrou equations 
as a set of first order equations using the osculating elements approach. 
To achieve this, we must first write the equations of motion in the form of a 
forced geodesic equation for a non-spinning particle: 
\begin{eqnarray}
\dot{u}^{\alpha}=\frac{d^2 x^{\alpha}}{ds^2} + \Gamma_{\rho\sigma}{}^{\alpha} 
\frac{d x^{\rho}}{ds} \frac{d x^{\sigma}}{ds} = f^{\alpha}, \label{osc1}
\end{eqnarray}
which we want to rewrite later in the form [\ref{Eq:OE}]. 
We denote the SSCs (i), (ii) and (iii) as ``T'' and ``F'' and ``w'', and consider first the ``T'' condition, $S^{ab}p_b = 0$.
In that case, we have $M=const$, but  $\dot{m}\left(u^{\alpha},\dot{u}^{\alpha} \right)=\dot{S}^{\alpha
\beta} \dot{u}_{\alpha} u_{\beta}$. We can 
introduce a new time variable, $\lambda$, with $\rmd\lambda=m\rmd\tau$ and use $\tilde{u}^{\alpha}$ to denote the coordinate velocity in the new coordinates $\tilde{u}^{\alpha} := \rmd x^{\alpha}/\rmd\lambda = u^{\alpha}/m$. The equations then become
\begin{eqnarray}
\frac{\rmd p^{\alpha}}{\rmd\lambda} + \Gamma_{\rho\sigma}{}^{\alpha} p^{\rho} \tilde{u}^{\sigma} 
= -\frac{1}{2} S^{\rho\sigma}\tilde{u}^{\mu} R_{\rho\sigma\mu}{}^{\alpha}, \nonumber \\
\frac{\rmd S^{\alpha\beta}}{\rmd\lambda} +\Gamma_{\rho\sigma}{}^{\alpha} S^{\beta\sigma} \tilde{u}^{\rho} + \Gamma_{\rho\sigma}{}^{\beta} S^{\alpha\sigma} \tilde{u}^{\rho} = 2 p^{[\alpha} \tilde{u}^{\beta]}, \nonumber \\
\tilde{u}^{\alpha} =\frac{\rmd x^{\alpha}}{\rmd\lambda} = \frac{1}{M^2}\left(p^{\alpha} + \frac{2 S ^{\alpha\beta} S^{\rho\sigma} R _{\beta\epsilon\rho\sigma} p^{\epsilon}}{4M^2 + S ^{\mu\beta} S^{\rho\sigma} R_{\mu\beta\rho\sigma}}\right), \label{Eq:Tu-p}
\end{eqnarray}
which now have no explicit dependence on $m$ and so we can proceed to write them in osculating element form. In particular, we can differentiate the third equation with respect to $\lambda$ and then use the first equation to get an equation for $\frac{\rmd \tilde{u}^{\alpha}}{\rmd\lambda} + \Gamma_{\rho\sigma}{}^{\alpha} \tilde{u}^{\beta} \tilde{u}^{\gamma}$ that depends only on position and velocity, and not on derivatives of $\tilde{u}^{\alpha}$. The explicit expression for the covariant total derivative of $\tilde{u}^a$ is given by:
\begin{eqnarray}
\frac{D\tilde{u}^{\alpha}}{d\lambda} &=&  m^2  \left( \tilde{u}^{\mu} +  \tilde{u}_{\beta} \frac{DS^{\mu\beta}}{d\lambda}\right) \frac{d}{d\lambda} H^{\alpha}{}_{\mu} + \Gamma_{\rho\sigma}{}^{\alpha} \tilde{u}^{\rho} \tilde{u}^{\sigma} \nonumber \\ 
&&- \Gamma_{\rho\sigma}{}^{\mu} \tilde{u}^{\rho} m^2 \left( \tilde{u}^{\sigma} + 
\tilde{u}_{\nu} \frac{DS^{\sigma\nu}}{d\lambda} \right) \left(H^{\alpha}{}_{\mu} - 
\frac{1}{M^2} \delta^{\alpha}_{\mu} \right) \nonumber \\
&&- \frac{1}{2} S^{\rho\sigma}  \tilde{u}^{\beta} R_{\rho\sigma\beta}{}^{\mu} \left(H^{\alpha}{}_{\mu} - 
\frac{1}{M^2} \delta^{\alpha}_{\mu} \right), \label{cov_new_u}
\end{eqnarray}
where we made use of the following abbreviation:
\begin{eqnarray}
H^{\alpha}{}_{\mu}:=\frac{2 S^{\alpha\beta} S^{\rho\sigma} R_{\rho\sigma\beta\mu}}
{4M^4+M^2 S^{\epsilon\lambda} S^{\kappa\nu} R_{\kappa\nu\epsilon\lambda}}. \label{def_H} \nonumber
\end{eqnarray}
The third equation in (\ref{Eq:Tu-p}) gives an implicit dependence of $p^{\alpha}$ on the spin tensor and 
velocity ($p^{\alpha} = p^{\alpha}(u^{\beta}, S^{\beta\gamma}))$ which we can use to integrate the (second) equation 
for the spin tensor.

We note, however, that the standard osculating element formulation of the equations implicitly imposes the condition that $\tilde{u}^{\alpha}\tilde{u}_{\alpha} = 1$ and hence $\tilde{u}^{\alpha}f_{\alpha} = 0$. This is no longer true after this change of variables. However, there is a way to put the equations into this standard osculating element form when there is an arbitrary force on the right hand side. To tackle this problem  we can again make a change of integration variable to a new variable, $q$ say. We then have
\begin{eqnarray}
\frac{\rmd x^{\alpha}}{\rmd\lambda} = \frac{\rmd q}{\rmd\lambda} \frac{\rmd x^{\alpha}}{\rmd q} \nonumber \\
\frac{\rmd^2x^{\alpha}}{\rmd\lambda^2} = \left(\frac{\rmd q}{\rmd\lambda}\right)^2  \frac{\rmd^2x^{\alpha}}{\rmd q^2} + \frac{\rmd^2q}{\rmd\lambda^2} \frac{\rmd x^{\alpha}}{\rmd q}
\end{eqnarray}
and the equations become
\begin{equation}
\frac{\rmd^2 x^{\alpha}}{\rmd\lambda^2} + \Gamma_{\rho\sigma}{}^{\alpha}\frac{\rmd x^{\rho}}{\rmd\lambda} \frac{\rmd x^{\sigma}}{\rmd\lambda} = f'^{\alpha} = \frac{1}{(\rmd q/\rmd\lambda)^2} \left(f^{\alpha} - \frac{\rmd^2q}{\rmd\lambda^2}\frac{\rmd x^{\alpha}}{\rmd q}\right) .
\end{equation}
We can impose the orthogonality condition be solving
\begin{equation}
 \frac{\rmd^2q}{\rmd\lambda^2} = \frac{f_{\alpha} \rmd x^{\alpha}/\rmd q}{g_{\alpha\beta}(\rmd x^{\alpha}/\rmd q)(\rmd x^{\beta}/\rmd q)}
\end{equation}
and the force becomes
\begin{equation}
f'^{\alpha} = \frac{1}{(\rmd q/\rmd\lambda)^2} \left(f^{\alpha} -\frac{f_{\gamma} \rmd x^{\gamma}/\rmd q}{g_{\mu\nu}(\rmd x^{\mu}/\rmd q)(\rmd x^{\nu}/\rmd q)} \frac{\rmd x^{\alpha}}{\rmd q}\right) .
\end{equation}
So, to compute the new force we need to know the value of $\rmd q/\rmd\lambda$. We can set this to one initially and then simultaneously integrate the equation
\begin{equation}
\frac{\rmd}{\rmd q} \left(\frac{\rmd q}{\rmd\lambda}\right) =  \frac{1}{\rmd q/\rmd\lambda} \frac{f_{\alpha} \rmd x^{\alpha}/\rmd q}{g_{\mu\nu}(\rmd x^{\mu}/\rmd q)(\rmd x^{\nu}/\rmd q)} .
\end{equation}
This is a somewhat complicated procedure, but the right hand sides of the new equations now do not depend on derivatives of velocity and so the problems identified above no longer apply.

The (iii) SSC ($w$-condition) is the most suitable for the osculating elements approach. In this case,
we use an arbitrary normalized time-like vector $w^{\alpha} w_{\alpha} = 1$  and impose
the following conditions
\begin{equation}
 w_{\alpha}S^{\alpha\beta} = 0, \;\;\; D_{\tau}w^{\alpha} = 0. 
\end{equation}
The vector field $w^{\alpha}$ is parallel propagated along the world line of the test mass and these conditions
imply $p^{\alpha} = m u^{\alpha}$ and $m$ is conserved. This SSC is the most suited for the osculating elements 
approach. 

Alternatively one can linearize the equations with respect to the spin $S_{\mu\nu}$
 In this case the relation between the velocity and the 4-momentum takes the simple form
\begin{eqnarray}
p^{\alpha}\stackrel{{\rm L}}{=} m u^{\alpha} \label{mom_vel_lin},
\end{eqnarray} 
and the supplementary conditions ``T'' and ``F'' coincide. 
The equations of motion  are now given by:
\begin{eqnarray}
\dot{u}^{\alpha} &\stackrel{{\rm L}}{=}& - \frac{1}{2m} S^{\rho\sigma} u^{\beta} R_{\rho\sigma\beta}{}^{\alpha}, \label{lin_mpd_1} \\
 \dot{S}^{\alpha\beta} &\stackrel{{\rm L}}{=}& 0.   \label{Eq:lin_mpd}
\end{eqnarray}
As is apparent from (\ref{Eq:lin_mpd}), this form of the equations of motion is suitable for the osculating orbits method, yielding a perturbing force of the form
\begin{eqnarray}
f^{\alpha} \stackrel{{\rm L}}{=} f^{\alpha}\left(u^{\alpha},S^{\alpha\beta}\right). \label{Eq:lin_force}
\end{eqnarray}

We will now stop considering a general background space time and 
focus on a particular choice: de Sitter. This is a spacetime 
with a constant curvature which is at the same time fully symmetric. 
This allows us to solve the equations of motion analytically and to gain 
better understanding of the trajectories and the role of the SSC.

The motion of spinning test particles  in de Sitter spacetime has previously been investigated by \cite{Obukhov:2010kn} where it was found that under the Tulczyjew SSC, the trajectory is a geodesic with the parallel transport of an appropriately defined spin vector. In addition, under the Frenkel-Pirani SSC, it was found that the trajectory is perturbed about a geodesic by an oscillatory motion but the final solution for the trajectory was left as a numerical integration. We focus on this oscillatory motion in more detail and relate it to motion under the $w$-condition.

The first Mathisson-Papapetrou equation \eqref{eq:MP1} simplifies in de Sitter spacetime to
\begin{equation}
D_{\tau} p^\alpha = \frac{1}{l^2}S^{\alpha\beta}u_\beta,
\end{equation}
where $l$ is a real constant, related to the Ricci scalar via $R = 12/l^2$. At first glance, it might appear that the Frenkel-Pirani SSC will lead to the simplest trajectories, as $D_{\tau} p^\alpha$ is identically zero in this case. However, due to the difference between 4-momentum and 4-velocity in \eqref{Eq:mom_vel}, this generically leads to non-geodesic motion.

We can write the equation of motion under both the Frenkel-Pirani and the $w$-condition in the same functional form, given by
\begin{eqnarray}
\label{eq:spineomgen}
D_{\tau} u^\alpha = \pm\frac{\omega \eta^{\alpha\beta\mu\nu}F_\beta u_\mu S_\nu}{\sqrt{\left(F_\sigma S^\sigma\right)^2+\left(F_\sigma F^\sigma\right)S^2}}, \quad D_{\tau} F^\alpha = 0, \quad D_{\tau} S^\alpha = 0,\nonumber\\
u_\alpha u^\alpha = 1, \quad F_\alpha S^\alpha = u_\alpha S^\alpha, \quad F_\alpha F^\alpha = u_\alpha F^\alpha, \quad S_\alpha S^\alpha = -S^2,
\end{eqnarray}
where $S^\alpha$ is a spin 4-vector constructed from the spin tensor such that the SSC is satisfied, $S$ and $\omega$ are real constants, and $\eta^{\alpha\beta\mu\nu}$ is the permutation symbol. Differentiating the equation for $D_{\tau} u^\alpha$, results in
\begin{equation}
D_{\tau}^2 u^\alpha = -\omega^2\left(u^\alpha - F^\alpha\right),
\end{equation}
demonstrating that $F^\alpha$ can be viewed as a forcing term for the oscillations.

The frequency of oscillation $\omega$ and the forcing term $F^\alpha$ are different for the two SSCs: for the Frenkel-Pirani case, we find
\begin{eqnarray}
F^\alpha &\stackrel{{\rm F}}{=}& \frac{m}{M^2} p^\alpha,\\
\omega &\stackrel{{\rm F}}{=}& \frac{2M}{S};
\end{eqnarray}
while under the w-condition,
\begin{eqnarray}
F^\alpha &\stackrel{{\rm w}}{=}& \left(u_\sigma w^\sigma\right)w^\alpha - \frac{u_\sigma S^\sigma}{S^2}S^\alpha,\\
\omega &\stackrel{{\rm w}}{=}& \frac{S}{2Ml^2}.
\end{eqnarray}

As we have an explicit equation for $D_{\tau} u^\alpha$, we could now numerically integrate, using the method of osculating elements, to find the trajectory. Instead, it is possible to find a general analytic solution to \eqref{eq:spineomgen} for the motion of spinning test particles in de Sitter spacetime. As a starting point, we note that the solution in Minkowski spacetime has been determined previously (see \cite{Kyrian:2007zz, Costa:2011zn, Kudryashova:2010gt}, for example). Under both the Tulczyjew and $w$-conditions, the particle follows a geodesic whilst under the Frenkel-Pirani condition, the particle undergoes purely circular motion, boosted along a central geodesic.

Since the de Sitter and Minkowski geometries are both maximally symmetric, it might be expected that a similar solution representing circular motion will be found in de Sitter spacetime. We are interested in the 16 components of the position $x^\alpha$, velocity $u^\alpha$, forcing term $F^\alpha$, and spin $S^\alpha$ 4-vectors, using spherically symmetric static coordinates. Using the ten isometries of the de Sitter spacetime and the four constraints in \eqref{eq:spineomgen}, it is possible to show that a completely general solution to the equations of motion is given by
\begin{eqnarray}
x^\mu(\tau)&=&\left\{t=u^t\tau,\,r,\,\theta=\frac{\pi}{2},\,\phi=u^\phi\tau\right\},\\
u^\mu(\tau)&=&\left\{u^t\!=\sqrt{\frac{1-r^2/l^2+\omega^2r^2}{1-2r^2/l^2}},\,u^r\!=0,\,u^\theta\!=0,\right. \nonumber \\
& & \left. u^{\phi}\!=\sqrt{\frac{r^2/l^2+\omega^2\left(l^2-r^2\right)}{l^2-2r^2}}\right\}, \\ 
F^\mu(\tau)&=&\left\{F^t\!=-\frac{u^\phi}{u^t}l^2F^\phi,\,F^r\!=0,\,F^\theta\!=0,\,F^\phi\!=-\frac{u^\phi\left(u^t\right)^2}{\omega^2l^2}\right\},\\
S^\mu(\tau)&=&\left\{S^t\!=-\frac{u^\phi}{u^t}l^2S^\phi,\,S^r\!=0,\,S^\theta\!=\pm\frac1{r}\sqrt{S^2+\frac{\omega^2l^4(S^\phi)^2}{(u^t)^2}},\,S^\phi\right\},\nonumber \\
\end{eqnarray}
where $r$ and $S^\phi$ are free constants. This solution explicitly corresponds to circular motion about the origin at a frequency that tends to $\omega$ in the limit that $l \rightarrow \infty$, consistent with the Minkowski result.

In spacetimes with fewer symmetries than de Sitter, we do not anticipate that such an exact analytic solution for the trajectory can be found, although progress can still be made. Different classes of pole-dipole orbits have been identified in the equatorial plane of Kerr~\cite{PhysRevD.90.064035} and it has been shown numerically that the motion of spinning test particles in Schwarzschild is of a helical nature~\cite{Plyatsko:2011gf}. The existence of the exact de Sitter solution can be used to further our understanding of spinning test particle trajectories in these more physical spacetimes.

In addition, the similarity of the solutions in de Sitter under the $w$-condition and the Frenkel-Pirani SSC will hopefully lead to a better understanding of these SSCs. Particularly, we note here that the product of covariant frequencies, $\omega_{\mathrm{P}}\,\omega_{\mathrm{w}}=1/l^2$ is dependent only on the curvature of de Sitter and not on the multipole moments of the test particle. If a similar fundamental link between the two SSCs exists in other spacetimes, it might allow us to infer properties of the Frenkel-Pirani trajectory by numerically integrating the simpler equation of motion under the $w$-condition.

\section{Detecting GW signals from EMRIs}
\label{EMRIsDetection}

In the previous two sections we have described the formation of EMRI systems and how the gravitational waves they generate can be modelled. 
Both those problems are very hard and not yet solved in full, and those astrophysical and 
theoretical uncertainties in EMRI rates and in models of the GW signal are coupled 
to the data analysis challenges. Before we describe specific data analysis algorithms 
for extracting EMRI GW signals from the detector data we will give a general description of the signal 
and the problems we face in data analysis. 

As mentioned earlier, an EMRI generates $10^5-10^6$ gravitational waveform cycles in the eLISA band. We therefore need to model it very accurately if we want to 
avoid systematic biases in the inferred parameter estimates. The expected signal-to-noise ratio (SNR) from those systems 
is not very high (probably less than 50), but during the  Mock LISA Data Challenges (MLDCs) successful extraction of EMRI signals with 
SNRs as low as 19 was demonstrated, using the same approximate EMRI model (the analytic kludge described earlier) for both injection and recovery. As described in Section~\ref{intro}, the EMRI signal depends on 14 parameters, if the spin of the CO is ignored, which is justifiable for mass ratios less than $\sim10^{-4}$.
It is convenient to describe the 
EMRI's dynamics in the frame fixed relative to the spin axis of the MBH. The spin direction is usually taken to be the $z-$axis, but we have full 
freedom in choosing the orientation of the $x,y$-axes, and this choice is degenerate with the initial azimuthal position of the CO.  Since the signal is long lived (stays in band for the entire duration of observation) there is a significant modulation of the amplitude and the phase of the waveform caused by the orbital motion of the 
detector. This allows us to measure the source sky position with a precision of a few degrees for signals with SNR$\sim20$ \cite{Babak:2009cj}. 

EMRIs are primarily GW sources for a future space-based detector like eLISA \cite{AmaroSeoane:2012km}, and
the data analysis discussion presented in this section is based on analysing data from such an instrument. 
Here we will always assume that the instrumental noise is Gaussian (but not white) and 
that EMRIs are the only GW sources in the data. These are not realistic assumptions for eLISA like data, but make the problem more tractable and the resulting algorithms are still likely to be effective when the assumptions are relaxed. For the purpose of developing 
data analysis algorithms and EMRI detection strategies we use somewhat simplified 
models of GW signal (in particular the analytic ``kludge'' model described in the 
subsection \ref{Waveforms}), which capture the main physical features present in the expected signal 
(periapsis and orbital precession, slow inspiral, Doppler modulation, multiple 
harmonics) and are also fast to generate numerically and so can be used for computationally expensive parameter estimation. The need to quickly evaluate hundreds of thousands of waveforms to perform data analysis is the main factor which 
prevents us from using more realistic models. If the data analysis algorithms do not use 
any model specific features, they can be easily ported to use the best GW signal model available at the time the data is analysed.

There are two data analysis challenges associated with the search for EMRI signals.
The first one is to find a signal in the noise, in other words to test the null hypothesis 
that the observed data is consistent with noise only. This could be a problem for 
signals with SNR below 20, however we do expect to see a few dozen signals from EMRIs 
with SNR above 20, which should be detected with high statistical significance. 
Therefore  we will concentrate on such reliably detectable signals. The situation will become 
more complex when other GW signals are present in the data (especially the foreground from Galactic 
white dwarf binaries)  and/or with realistic instrumental noise. We do expect 
some environmental and instrumental artefacts to be present in the data and the LISA Pathfinder 
\cite{Antonucci:2012zz} measurements (scheduled for launch in July 2015) will allow us to simulate a more realistic 
eLISA data stream in the near future.   

The second problem is what we will focus on in the rest of the current section. The large dimensionality of the parameter space of possible signals makes a grid-type search completely infeasible, so instead we will rely 
on (pseudo)-stochastic search methods, primarily based on Markov chain Monte-Carlo 
(MCMC) techniques. Various implementations of MCMC for searches for EMRIs signals
are described in \cite{Babak:2009ua, Babak:2009cj, Wang:2012xh}, but the basic idea is to construct a chain which moves 
predominantly in the direction of increasing likelihood. The complication is that the EMRI likelihood 
hyper-surface has numerous local maxima some of which could be as much as $70-80\%$ of the 
global maximum and these local maxima are widely separated in the parameter space. 
The problem is similar to finding the tallest tree in a forest. A standard MCMC
based search will reach a local maximum and get stuck there for a significant  number 
 of steps. Theoretically  MCMC has a non-zero probability of exploring the whole parameter space 
 and finding the global maximum, but in practice it can get stuck on a strong local maximum 
 for a very long time. Since we consider here only clearly detectable signals, when we refer to a
 detection we will mean successfully finding the global maximum of the likelihood (which is near the true parameters 
 of a simulated signal and by ``near'' we mean comparable to the expected statistical deviations due to the presence of detector noise).
 
 In order to detect a GW signal from an EMRI we need an algorithm which can explore efficiently 
 a large part of the parameter space and at the same time concentrate more on regions 
 of high likelihood. Parallel tempering MCMC is one such algorithm and it was used 
 in the MLDCs by N. Cornish \cite{Cornish:2008zd}. Here we describe two other methods which share 
 the same core principle, based on understanding and exploiting the reason for the presence of local 
 maxima. To understand this reason, we need to look carefully at the GW signal.
 The GW signal from an EMRI is a superposition of harmonics of three fundamental frequencies, which slowly evolve as the CO inspirals. 
 \begin{eqnarray}
 h(t) = \sum_{l,m,n} h_{lmn}(t) = 
 \Re \left(  \sum_{l,m,n} A_{lmn}(t) e^{i(l\Phi_r + m\Phi_{\theta} +n\Phi_{\phi})} \right)
 \end{eqnarray}
 
\begin{figure}
\begin{center}
\includegraphics[height=8cm]{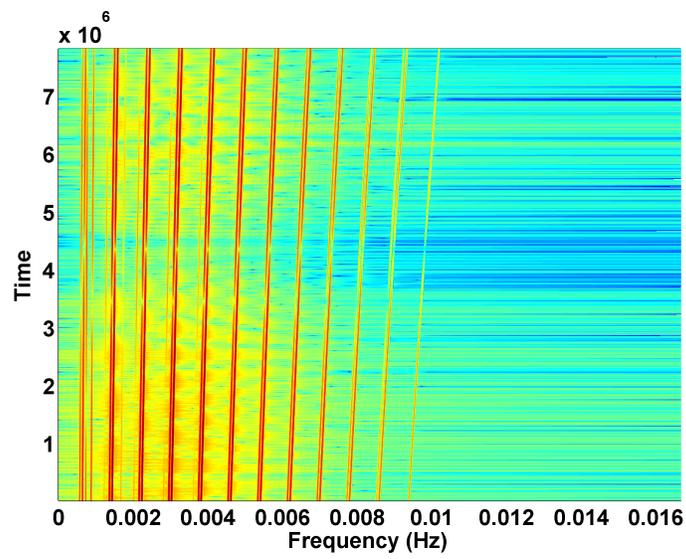}
\end{center}
\caption{The time-frequency plot of a typical GW signal from an EMRI, there are 30 clearly identifiable  
harmonics slowly evolving in time. The amplitude is colour coded. The time is in seconds. }
\label{F:EMRI_tf}
\end{figure}

 These fundamental frequencies (instantaneously or 
 for a geodesic motion) are associated with three degrees of freedom: the radial frequency 
 is associated with eccentric motion from periapsis to  apoapsis and back; the polar frequency 
 ($\theta-$motion) is associated with spin-orbital coupling and the resulting  precession of the orbital 
 plane around the spin axis of the MBH; and finally the frequency 
 of azimuthal motion \cite{Schmidt:2002qk, Drasco:2005kz}. The frequencies evolve under 
 radiation reaction (self-force) on a time scale associated with the mass ratio, which is 
 for EMRIs significantly longer than the orbital time scale. As the CO spirals toward the MBH 
 the overall amplitude of the signal is slightly increasing but the amplitude of individual 
 harmonic depends on the instantaneous orbital parameters like eccentricity and 
 inclination. Due to orbital circularisation under radiation reaction \cite{Peters:1963ux} the 
 amplitude of some harmonics (high $l$) will decrease while that of some other (low $l$) harmonics will increase, 
 but in all cases the amplitude of each harmonic is a smooth and slowly varying function
 of time. We can construct a periodogram of the EMRI signal, and it looks like a comb 
 in the time-frequency plane, see Figure~\ref{F:EMRI_tf} as an example.

The global maximum corresponds to the case when two combs representing a signal and a search template 
coincide exactly in amplitude everywhere in the time-frequency plane. The reason for the local maxima is a partial overlap between the signal harmonics and the harmonics of a template. 
These might not be the same harmonics (the same set of $l,m,n$) and the strength of a given local 
maximum will depend on how long (in frequency and in time) the harmonics of the signal and template 
coincide.

In the search for a GW signal we use matched filtering 
which is an optimal detection technique in the presence of Gaussian noise and can be seen as an inner 
product of the data $x(t) = n(t) + s(t)$ with a template $h(t)$. Here $n(t)$ is the instrumental  
noise and the signal $s(t) = s(t; \vec{\lambda})$  depends on the parameters of the source 
($\vec{\lambda}$), which we are trying to estimate. The inner product is defined as 

\begin{equation}
(x, h) = 2\Re \int_{0}^{\infty}\frac{\tilde{x}^* (f)\tilde{h}(f)}{S_n(f)} df, 
\label{inprod_f}
\end{equation}
where tilde denotes a Fourier transformed quantity and $S_n(f)$ is the one-sided 
power spectral density of the noise in the detector. If the signal is confined within a narrow frequency band 
around $f_0$, so that we can treat $S_n(f_0)$ as almost constant, the inner product 
can also be written in the time domain in a simple form:
\begin{equation}
(x,h) \approx \frac1{S_n(f_0)} \int_0^T x(t)h(t)dt,
\label{inprod_t}
\end{equation} 
where $T$ is the observation time (or duration of a template).  The assumption that 
$S_n(f)$ is approximately constant over the signal evolution is valid for signals of duration up to
2-5 months (dependent on the parameters). Since the amplitude of an EMRI signal is 
a slowly growing function of time,  one can see from Eq.~(\ref{inprod_t}) that the SNR
($SNR^2 = (s, s)$) roughly grows as the square root of the observation time.  
We can use a maximum likelihood estimator to determine the GW parameters. The likelihood ratio  is given by 
\begin{equation}
\Lambda(\vec{\lambda}) = \frac{P(x|h(\vec{\lambda})}{P(x|0)} = 
e^{(x,h(\vec{\lambda})) - \frac1{2}(h(\vec{\lambda}), h(\vec{\lambda}))},
\end{equation} 
where $P(x|h(\vec{\lambda})$ is the probability that the data $x$ would be observed when a signal corresponding to the specified set of parameters is present in the 
data and $P(x|0)$ is the probability that the data would be observed when no signal was present.
Usually the likelihood (or log-likelihood) can be maximised over some parameters of the 
signal analytically, whereas maximisation over other parameters requires a numerical search. 
The analytically  maximised likelihood is quite often referred as the $F$-statistic 
\cite{Jaranowski:1998qm, Wang:2012xh, Babak:2009ua}. 

\begin{figure}
\begin{center}
\includegraphics[height=4.5cm]{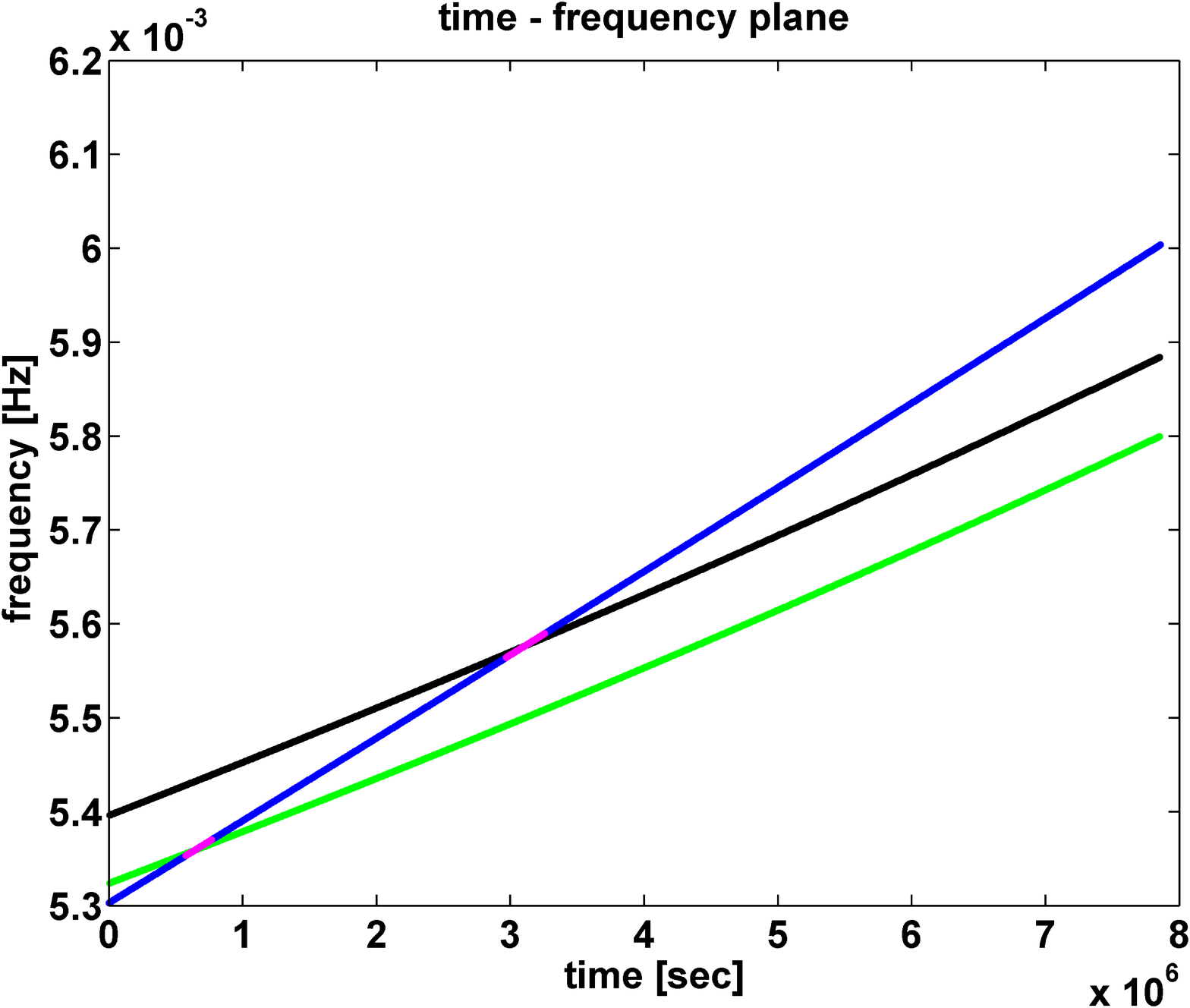}
\includegraphics[height=4.5cm]{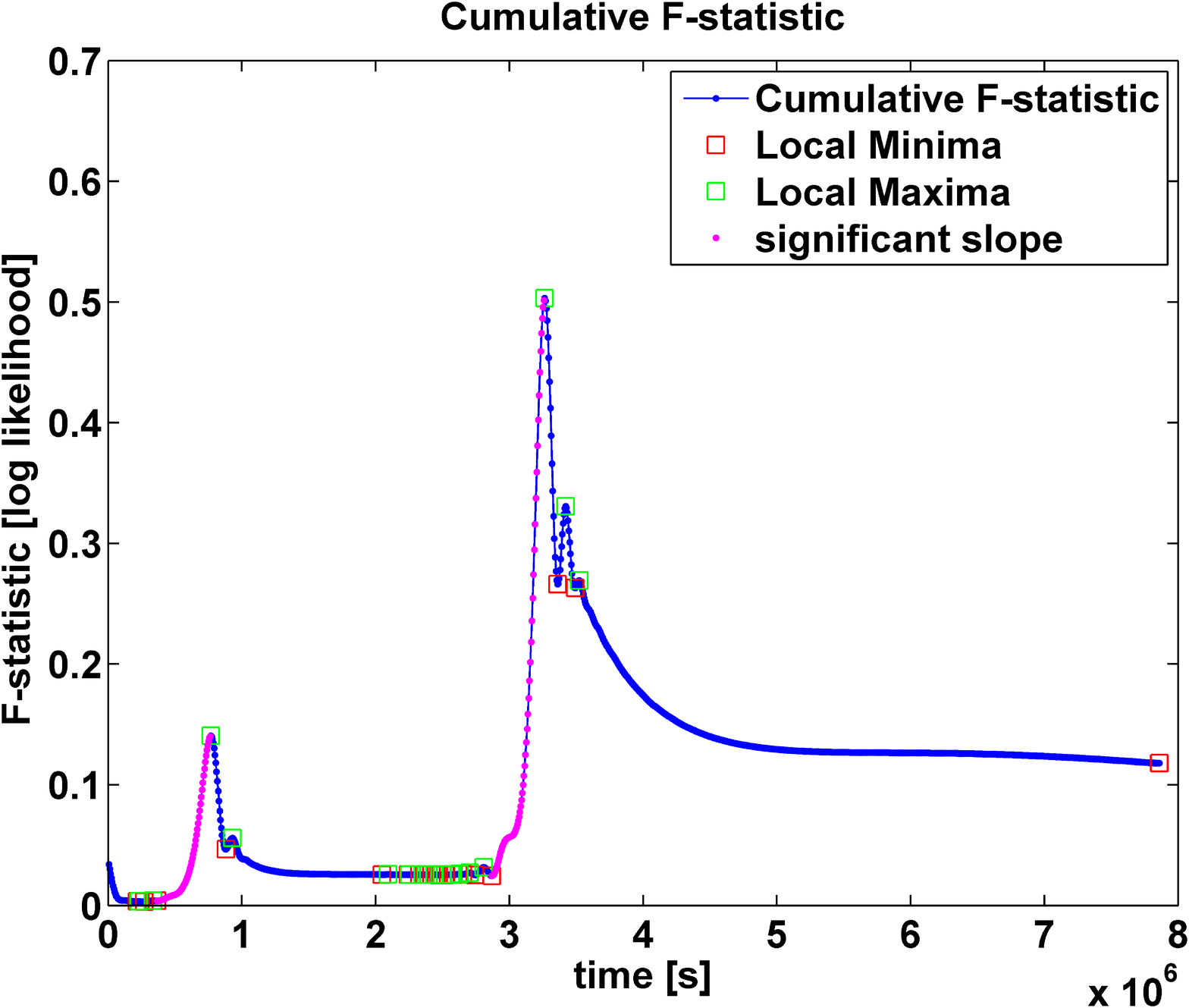}
\end{center}
\caption{Cartoon showing two harmonics of a signal in green and black (black being stronger)
and a harmonic of the template intersecting the signal at two instances (left plot). In the right plot 
we give the corresponding accumulation of the $F$-statistic in time. The two significant positive slopes 
(in pink) corresponds to two instances of overlaps between a signal and a template.}
\label{F:cartoonHarm}
\end{figure}

Based on the equations (\ref{inprod_f}), (\ref{inprod_t}) we can introduce a cumulative 
likelihood (or cumulative $F$-statistic) in the time and/or in the frequency domain by varying the 
upper limit of integration. If the template matches the signal exactly we expect to have 
steady growth of the cumulative $F$-statistic as a function of time or frequency
(in other words it should be a monotonic and not decreasing function). In the case of a
local maxima we will observe ``bursts'' of increase in the $F$-statistic around instances 
of time (or frequency) where one or more harmonics of the template and signal match. 
This is illustrated in~Figure \ref{F:cartoonHarm}, the left panel shows schematically a harmonic of a template successively intersecting and overlapping with two different harmonics of the signal, one of which (in black) if stronger than the other. 

In the right panel of the same Figure, we show the corresponding  accumulation of the $F$-statistic, and 
the instances of two intersections are clearly seen here as a rapid increase in the 
$F$-statistic. This illustrates nicely the reason for the presence of strong local maxima in the parameter space which we hit 
while constructing the Markov chain: harmonics of a signal can reproduce (overlap) 
one or a few strong harmonics of a signal for a span of time sufficient to accumulate 
a significant value of the detection statistic. This makes a ``curse'' into a ``blessing'': 
we can use the information of the locations of the local maxima to guide the search to find 
the global maximum of the likelihood. This is a key part of the search for EMRIs 
and the main basis for the two specific methods described in the following subsections.
We find many local maxima by running multiple MCMC chains with different seeds, and then analyse the 
accumulation of the $F$-statistic to identify the parts (harmonics) of the signal that were found at each of 
those  local maxima. Then we use this information to run a constrained 
MCMC (as described in subsection \ref{S:cMCMC}) or place them on the time-frequency 
plane and fit them with the harmonic tracks of  a template by varying the source
parameters (as described in subsection \ref{S:phenEMRI}).


\subsection{Constrained Markov Chain Monte Carlo search}
\label{S:cMCMC}
In this subsection we will summarize the method which was successfully used to analyse the Mock LISA 
data  challenge \cite{Babak:2009cj} and described in greater detail in \cite{Babak:2009ua}.
In this method we  split the data into 6-month long subsets and start by analysing each of them
separately, before joining them together once we have started to lock onto the signal. 

In the first step we perform a stochastic search: we randomly draw parameters  from the prior 
range and evaluate their likelihoods. This is continued until multiple statistically significant points have been identified in the parameter space. 
Those points are then refined by running small MCMC chains seeded at those points. The local maxima 
are then analysed to find common harmonics (in time and frequency). These are identified as sections of harmonics of the true signal, although usually
we do not know the associated harmonic indices.

In the second step we run a constrained MCMC. The sections of harmonics found in the first stage  
serve as constraints. We do realise that those constraints might not be exact, so we first run the
MCMC with the frequency constraints and adjust the other parameters then we
release the constraint and allow the code to adjust the constrained frequencies before fixing these again and repeating. 
This works very well in practice, even if the frequency of some of the (especially weak) harmonics 
was not determined very accurately initially. We also run several chains simultaneously to check for
convergence to a global maximum. 

In the third step we join the 6-month-long subsets of data together and let the chains adjust to match together the best 
found solutions in each subset.  This method was used to analyse simulated data with a 
single relatively strong (SNR between 50 and 130) EMRI signal (\cite{Babak:2009ua}). The identification
of a signal was remarkably good with an ultra-precise recovery of the system parameters. The technique was also used to analyse 
the third Mock LISA data challenge data set, for which there was a single data set with five weak 
(SNR about 20) EMRI signals. The technique successfully identified two signals, while for the other three signals
we identified that they were present but did not determine reliable estimates of their parameters before the challenge deadline.

\subsection{Detection of EMRIs using a phenomenological template family}
\label{S:phenEMRI}

In this subsection we summarise the method described in \cite{Wang:2012xh}.
The main idea of this approach was to detect GW signals from EMRIs in a model independent way using 
a minimal set of assumptions about the signal: (1) the orbital motion can be described by six slowly 
(on the radiation reaction time scale) changing quantities; and (2) the signal is represented by a set 
of harmonics  of those (three)  orbital frequencies with slowly changing amplitude. Those are rather 
mild constraints and should describe also ``dirty'' EMRIs where the orbital motion is perturbed 
either by the astrophysical environment or by a deviation in the spacetime geometry of the central BH \cite{Babak:2010ej, Barausse:2014tra}. 

We can use the assumption of slow frequency and amplitude evolution to decompose the phase and 
amplitude of each harmonic as a Taylor series and perform the search over the 
coefficients of the Taylor expansion.  We call this a phenomenological EMRI template --- the  
relationship between the Taylor series coefficients and the physical parameters depends on the specific model 
for the GW signal from an EMRI system.  By searching over phenomenological parameters (Taylor
coefficients) we do not restrict ourselves to any specific model within the framework of our 
assumptions above. The truncation of the Taylor series and the number of harmonics included depends 
mainly on the SNR of the signal: for weak signals we have to use a higher order expansion 
in order to match the signal for a longer time. Detection of EMRI signals in this model 
independent way allows us to relax stringent requirements on the accuracy of the theoretical 
model and to test alternatives to the assumption of a CO inspiral occurring in a pure vacuum Kerr spacetime.

\begin{figure}
\begin{center}
\includegraphics[height=8cm]{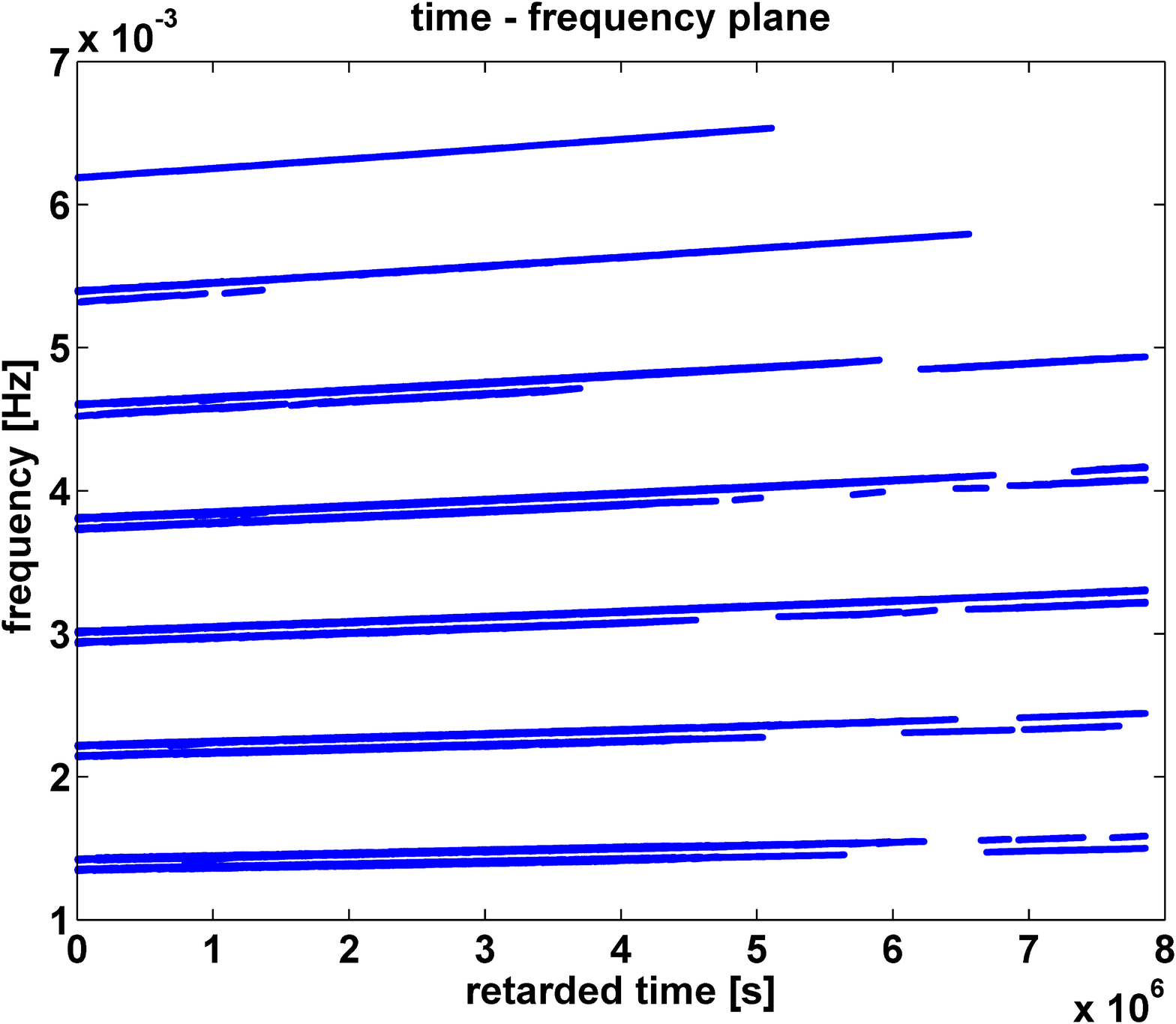}
\end{center}
\caption{Recovered patches of the signal corresponding to a strong accumulation of the 
$F$-statistic.}
\label{F:tfmap}
\end{figure}

Here we describe the simulations performed in 
\cite{Wang:2012xh}. Three month long data sets were simulated containing an EMRI signal ($SNR=50$)
using the numerical kludge as a model. Multiple MCMC searches using the phenomenological 
templates were carried out with different starting seeds. The results were collected and analysed for the presence of 
local maxima. For each identified maximum a patch of the signal harmonic which was found 
was extracted and placed on the time-frequency plane. The resulting map looks as presented in 
Figure \ref{F:tfmap}.
In this example the injected source was a strong signal and the method recovered 13 harmonics. In more realistic 
cases we would expect to recover 3-5 harmonics only. We note that the strong harmonics (at low frequency)
are better recovered (through the full duration of the observation). Notice also that the last month of the data 
is recovered less well than the first two, which is due to the orbital motion of the detector --- the antenna beam pattern 
during the last month is pointing away from the source.

In the second stage it is necessary to assume a certain EMRI model, so that the found harmonics can be identified and  the physical parameters of the system recovered. In particular it is here that we can assume several alternatives: a CO spiralling toward a Kerr MBH, a CO spiralling into a massive boson star, a ``dirty'' Kerr black hole (a bumpy BH 
or a complex astrophysical environment).  Once the model is assumed, we can find the set of parameters which
give the  best fit to the found set of harmonics (in amplitude and in their evolution). One can use a simple chi-square test of goodness of fit to estimate how well the assumed model describes the observed harmonic tracks and hence make a statement about the model. Results for the recovery of orbital parameters if the same model is used for recovery and signal generation were presented in \cite{Wang:2012xh}.

\section{Conclusion}

In this article we have described one of the most interesting GW sources for 
the future space based gravitational wave observatory eLISA. We have briefly described the various channels for EMRI formation 
and expected event rates. Then we went through an inventory of available models for the GW signal 
generated by EMRIs. We also briefly discussed the osculating element approach 
for integration of the forced (under radiation reaction) motion of a CO in Kerr spacetime, and its 
application to the case of a spinning CO. One non-trivial question is the influence of the 
spin supplementary condition on the computed motion of a spinning  CO and we have addressed this by looking at a 
simplified case: the motion of a spinning test mass in de Sitter spacetime. This should provide guidance on 
how to proceed in the case of a Schwarzschild or Kerr spacetime. Finally we have described the challenges which we will 
face in extracting GW signals generated by EMRIs from eLISA data. The main problem 
is to search for a global maximum of likelihood in the multidimensional parameter space, when 
multiple strong local maxima are also present. We have described how one can extract useful information about the 
signal from the locations of those local maxima in order to direct the search to the correct solution. In addition we have outlined
the possibility that these methods can be used to verify that the central massive compact object is indeed described by 
the Kerr metric, as predicted by general relativity.


\bibliographystyle{unsrt}

\bibliography{Babak_eom_proceedings_2013}

\end{document}